\documentstyle[amssymb,aps,prb,preprint]{revtex}
\begin{document}
\title{Breakdown of the lattice polaron picture in ${\rm {\bf La_{0.7}Ca_{0.3}MnO_3}%
}$ single crystals}
\author{S. H. Chun$^1$\cite{byline}, M. B. Salamon$^1$, Y. Tomioka$^2$, and Y. Tokura%
$^{2,3}$}
\address{$^1$ Department of Physics and Materials Research Laboratory, University of\\
Illinois at Urbana-Champaign, Urbana, Illinois 61801-3080, USA\\
$^2$ Joint Research Center for Atomic Technology (JRCAT), Tsukuba 305, Japan%
\\
$^3$ Department of Applied Physics, University of Tokyo, Tokyo 113, Japan}
\date{\today}
\maketitle

\begin{abstract}
When heated through the magnetic transition at $T_C$, La$_{0.7}$Ca$_{0.3}$MnO%
$_3$ changes from a band metal to a polaronic insulator. The Hall constant $%
R_H$, through its activated behavior and sign anomaly, provides key evidence
for polaronic behavior. We use $R_H$ and the Hall mobility to demonstrate
the breakdown of the polaron phase. Above $1.4T_C$, the polaron picture
holds in detail, while below, the activation energies of both $R_H$ and the
mobility deviate strongly from their polaronic values. These changes reflect
the presence of metallic, ferromagnetic fluctuations, in the volume of which
the Hall effect develops additional contributions tied to quantal phases.
\end{abstract}


It is now well established that the insulator-metal transition and
accompanying colossal magnetoresistive effect (CMR) in doped LaMnO$_{3}$
arise as small lattice polarons, stable at high temperature, give way to
band conduction at low temperature\cite{Ramirez}. The process by which this
occurs is, of course, the key to understanding this and many similar
materials in which charge, lattice, orbital, and spin degrees of freedom are
intimately linked. The breakdown of the band structure is readily identified
with the temperature at which a rapid increase in resistance and the onset
of strong negative magnetoresistance arises in well-prepared single crystal
samples. Less evident is the temperature at which the small-polaron picture
ceases to be valid and, indeed, the nature of charge carriers in the
transition region between the two. Recently De Teresa et al. detected the
presence of small ferromagnetic clusters below $1.8T_{C}$\cite{deteresa}.
However, the way they participate in the macroscopic transport properties
remains poorly understood. A possible picture, based on resistivity and
thermopower data, suggests that the two phases, polaronic insulator and band
metal, coexist and that the metallic concentration increases as magnetic
order sets in, resulting in a percolation transition\cite{2fluid}.

A key piece of evidence for the high temperature polaronic phase comes from
Hall effect measurements. At high temperature ($>1.4T_{C})$ previous results 
\cite{Jaime} found the Hall constant to be negative, despite the hole doping
of the material, with the Hall constant $R_{H}=\rho _{xy}(B)/B$ showing
activated behavior with characteristic energy $E_{H}\approx \frac{2}{3}%
E_{\sigma },$ where $E_{\sigma }$ is the activation energy for ordinary
conductivity $\sigma _{xx}$. These results are in accord with the theory of
adiabatic hopping of small polarons\cite{EH}. It is natural, therefore, to
seek evidence for the breakdown of the small-polaron phase in the behavior
of the Hall resistivity $\rho _{xy}(B)$ as the temperature approaches the
simultaneous insulator-metal/ferromagnetic transition. As we will show, the
activation energy changes abruptly at $1.4T_{C}$ from the polaronic value of 
$\frac{2}{3}E_{\sigma }$ to a much larger value, $1.7E_{\sigma }$, clearly
marking the breakdown of the lattice polaron picture in a transport
property. In fact, the effective activation energy of the conductivity
begins to decrease from $E_{\sigma }$ at roughly the same temperature,
making the discrepancy even greater. Yet more dramatically, the product of
the Hall mobility and temperature $\mu _{H}T=-\sigma _{xx}R_{H}T,$ which
should decrease monotonically with decreasing temperature, in fact exhibits
a minimum at this same cross over temperature. We have argued elsewhere that
Hall effect near $T_{C}$ arises from the combined effect of the double
exchange mechanism, in which the spin of the charge carrier must follow the
local magnetization, and the partial, but nonuniform, magnetic order that
arises near the ferromagnetic transition\cite{Chun}. Well below $T_{C}$,
where metallic conductivity dominates, the Hall effect resembles that
expected for a metallic ferromagnet.

To carry out this study, we have measured the longitudinal and transverse
(Hall) resistivity and the magnetization of La$_{0.7}$Ca$_{0.3}$MnO$_{3}$
single crystal samples ($T_{C}=216.2$ K), spanning the full temperature
range, from metallic ferromagnet to insulating paramagnet. High quality
single crystals of La$_{0.7}$Ca$_{0.3}$MnO$_{3}$ were prepared by the
floating-zone method. Electron Probe Micro Analysis revealed a slight
deficiency of Ca by 0.01, which partially explains the lower transition
temperature compared to polycrystalline samples; details of the growth
conditions can be found elsewhere\cite{Sample}. The specimen used in this
experiment has a bar shape with dimensions of $3\times 1\times 0.24$ mm$^{3}$%
. Contact pads for Hall resistivity measurements were made by sputtering $%
\approx $1500 \AA\ of gold\ through a mask. Gold wires were then attached
using slowly drying silver paints. The contact resistances after annealing
were about 1 $\Omega $ at room temperature. We adopted both a low frequency
ac and a dc method for the measurements. The transverse voltage signal was
first nulled at zero field at each temperature by a potentiometer and the
change in the transverse voltage recorded as $H$ was swept from +7 T to -7 T
and back for averaging. Following the transport measurements, the
magnetization was measured by a 7 T SQUID magnetometer on the same sample.

The longitudinal resistivity of our crystal is shown as a function of
temperature in the inset of Fig.~1 for two different fields. The small
resistivity at low temperature ($\rho _{xx}^0\approx $ 140 $\mu \Omega $cm
at 10 K) and the sharp insulator-metal transition indicate the high quality
of the crystal used in this experiment. We tentatively divide the
temperature into three different regions according to the behavior of $\rho
_{xx}$. In region I (metallic region, below 200 K) $\rho _{xx}$ shows
metallic behavior; a term proportional to $T^2$ dominates the temperature
dependence and has been ascribed to electron-electron\cite{e-e} or
one-magnon scattering processes\cite{magnon}. The magnetoresistance (MR),
defined as $\left[ \rho _{xx}(H=0)-\rho _{xx}(H=7\text{ T})\right] /\rho
_{xx}(H=7$ T$)$, is negligible in this region. In region II (transition
region, between 200 K and 300 K) $\rho _{xx}$ increases sharply and shows a
maximum at 230 K. The insulator-metal transition temperature $T_{IM}$,
defined by the peak in $d\rho _{xx}/dT$, is 222.5 K. Although this
transition is accompanied by paramagnet-to-ferromagnet transition, $T_{IM}$
is a few degrees higher than the Curie temperature $T_C$ = 216.2 K,
determined by scaling analysis of high field dc magnetization. The
application of high magnetic field reduces $\rho _{xx}$ drastically, thus
producing CMR. The maximum reduction occurs at 230 K, the resistivity peak
temperature, where $\rho _{xx}$ decreases to less than 4\% of its zero field
value under 7 T (MR = 2600 \%).

The high temperature behavior of $\rho _{xx}$ is well described by small
polaron hopping theory in region III (polaron region, above 300 K).{\em \ }%
In the adiabatic regime, the conductivity is given by 
\begin{equation}
\sigma _{xx}\equiv \frac 1{\rho _{xx}}=\sigma _0\exp (-E_\sigma /k_BT),
\label{eq-rho}
\end{equation}
and $\sigma _0=g_de^2\nu _0\delta /ak_BT$. Here, $\nu _0$ and $a$ are the
attempt frequency and hopping distance, respectively. We take $a$ to be the
Mn-Mn spacing, 0.39 nm. The factor $g_d$ is determined by hopping geometry
and $\delta $ is the carrier concentration per Mn site\cite{EH}. The large
value of our prefactor $\sigma _0\approx 4000$ $\Omega ^{-1}$cm$^{-1}$ at
300 K confirms the applicability of the adiabatic limit. From the fit to the
zero-field data, we have $E_\sigma =$106 meV, a typical value in this system 
\cite{Jaime}. The characteristic frequency $\nu _0$ is 1.5$\times $10$^{13}$
Hz if we use $g_d=5$ and $\delta =0.3$, and is close to optical phonon
frequencies of Mn-O-Mn bond bending (1.4$\times $10$^{13}$ Hz) and
stretching (1.9$\times $10$^{13}$ Hz)\cite{Yoon}. As the temperature is
reduced, the data deviate from Eq.~(\ref{eq-rho}) and large
magnetoresistivity sets in. We concentrate in the remainder of this paper on
the breakdown of small-polaron dynamics at the boundary between regions II\
and III.

The main panel of Fig.~1 shows the field dependence of $\rho _{xy}$ at
several temperatures. The general behavior shares the common features found
in other crystals\cite{Asamitsu,ChunLPMO} and thin films\cite{Matl,Jacob}.
At low temperatures (region I) the behavior is not dissimilar to a band
ferromagnet: $\rho _{xy}$ decreases initially due to the anomalous Hall
effect (AHE) and then increases above 1 T due to the ordinary Hall effect
(OHE) of hole-doped material. In region II (transition region), $\rho _{xy}$
is a strongly non-linear function of field. Finally, in region III, $\rho
_{xy}$ is again linear in $B$, but the slope is now negative.

The Hall effect due to the adiabatic hopping of small polarons was first
treated by Emin and Holstein\cite{EH}. In that model, the Hall effect arises
from interference between direct and indirect hops between two sites via
neighboring sites. The trajectory around visited sites encloses magnetic
flux, and the interference arises from the Aharonov-Bohm phase thereby
induced. When that trajectory involves an odd number of sites, the Hall
effect is always negative. Indeed, in region III, the low-field limit of the
Hall coefficient $R_H\equiv \rho _{xy}/B$ is negative, an effect observed
previously by Jaime et al.\cite{Jaime} on film samples with lower transition
temperatures. It was argued in that paper that both the magnitude of $\sigma
_0$ and the sign of the Hall effect point to direct Mn-Mn hops and, as a
result, to three-site hopping trajectories, and to the introduction of the
geometrical factor $g_d$ in the expression for $\sigma _0$. Extending the
Emin-Holstein model to this geometry leads to the prediction, 
\begin{equation}
R_H=-\frac{g_H}{g_d}\left( \frac{F(J/k_BT)}{ne}\right) \exp \left( \frac{E_H%
}{k_BT}\right) ,  \label{eq-RH}
\end{equation}
where $E_H=(2E_\sigma +E_s-J)/3;$ $J$ is the magnitude of the transfer
integral; $E_s,$ the chemical potential of polarons; and $F,$ a
dimensionless function of temperature. We set the geometrical factor $%
g_H=2/5 $\cite{Jaime}, and estimate $J-E_s$ from the band width and
thermopower to be $\lesssim 20$ meV. As we can see in Fig.~2, $R_H(T)$
agrees well with the small polaron theory in for temperatures above 300 K.
The slope gives $E_H=73\pm 7$ meV as compared with $(2E_\sigma +E_s-J)/3\geq
65$ meV. The lower curve in Fig.~2 shows that $E_\sigma $ is constant over
the same temperature range (note the inverted scale). From the prefactor in
Eq.~(\ref{eq-RH}) we estimate a carrier density $n\simeq 2.3\sim 3.7\times $%
10$^{27}$ m$^{-3}$ (depending on the magnitude of $J$), close to the nominal
doping level 5.0$\times $10$^{27}$ m$^{-3}$.

The Hall coefficient $\left| R_H\right| $ increases more rapidly below 300 K
with an effective activation energy $\simeq 1.7E_\sigma ,$ while the
effective activation energy of the longitudinal conductivity decreases. This
deviation from the small polaron theory can be seen even more clearly in the
Hall mobility $\mu _H\equiv R_H/\rho _{xx}$ which measures the excess jump
rate induced by the applied field. We expect $\mu _HT$ to decrease
monotonically as the temperature is reduced, with an activation energy $%
E_\mu \sim (E_\sigma -E_s+J)/3\leqslant 41$ meV. As shown in the inset of
Fig.~2, $\mu _HT$ decreases as the system cools to $T=$ 300 K, with
activation energy $E_\mu \approx 33$ meV, but then {\it increases} on
further cooling. Note that $E_H+E_\mu =E_\sigma ,$ as predicted by
small-polaron theory. Another advantage of using $\mu _H$ comes from the
fact that it is almost field-independent above $T_C$\cite{Matl,ChunJAP}, so
that the crossover behavior around 300 K is not subject to a low field
constraint. The simultaneous decrease in the effective activation energy $%
E_\sigma $ and increase in field-induced hopping measured by the Hall
mobility points to a fundamental change in carrier behavior. Within the
double-exchange model, hole motion is enhanced when neighboring core spins
are more nearly aligned. This suggests that local magnetic fluctuations,
which grow and become slower with decreasing temperature, lead to more rapid
carrier motion within the fluctuation volume. As a consequence, the material
enters a two-phase regime in which more conducting regions accompany
magnetic fluctuations and at the same time open a new channel for Hall
mobility arising from the constraint that the hopping carrier follow the
local spin texture. In this picture, small-polaron behavior disappears only
when spin correlations increase to the point that they dominate lattice
effects, with the features observed near $1.4T_C$ marking that boundary.

Below $T_{C},$ the sample becomes increasingly metallic, and we can discuss
the Hall effect in the terms usually used for ferromagnetic metals\cite{Hurd}%
, namely as a sum of ordinary [$R_{0}\left( T\right) ]$ and anomalous [$%
R_{S}\left( T\right) ]$ Hall contributions 
\begin{equation}
\rho _{xy}\left( B,T\right) =R_{0}\left( T\right) B+\mu _{0}R_{S}\left(
T\right) M\left( B,T\right) ,  \label{eq-AHE}
\end{equation}
At 10 K, we find $R_{0}$ = 2.3$\times $10$^{-10}$ m$^{3}$/C which, if we
assume the free electron model, corresponds to an effective charge carrier
density $n_{eff}=1/eR_{0}=$2.7$\times $10$^{28}$ m$^{-3}$ or 1.6 holes/Mn,
much larger than the nominal doping level (0.3 holes/Mn). Large effective
carrier densities have been observed commonly in other crystals and thin
films\cite{Asamitsu,ChunLPMO,Matl,Jacob} and can be explained when details
of the Fermi surface are taken into account\cite{ChunLPMO}. Because $R_{0}$
is constant in region I, we can easily subtract it from the data to obtain $%
R_{S}.$ In regions II and III, the separation between $R_{0}$ and $R_{S}$ is
less clear. However, a fit to Eq. 3 in similar materials showed that $R_{S}$
was more than an order of magnitude larger than $R_{0}$ near $T_{C}$\cite
{ChunLPMO}. Moreover, $\mu _{0}M$ is much larger than $B$ at low fields
through region II and the initial part of region III. Thus, in those
regions, we take the low-field values of $\rho _{xy}/\mu _{0}M$ as $R_{S}$
and superpose a plot of $R_{S},$ normalized to its peak value, and the
longitudinal resistivity $\rho _{xx},$ similarly normalized in Fig. 3. The
inset shows the data on a linear plot, and it appears that $R_{S}$ is
proportional to $\rho _{xx},$ and that the anomalous Hall effect is
dominated by skew-scattering\cite{Smit} in region I. However, when plotted
on a logarithmic scale in the main panel, it is clear that there are strong
deviations, and that the anomalous Hall effect in region I is proportional
to $\rho _{xx}^{2},$ which is the signature of side-jump processes\cite
{Berger}. While it may appear that $R_{S}$ is proportional to $\rho _{xx}$
in region II, we will argue elsewhere that this regime is dominated by the
effect of a Berry-like (Pancharatnam) phase induced by the Hund's rule
constraint that the hopping charge carrier follow the local magnetic texture 
\cite{Chun}. The deviation in region III we attribute, of course, to the OHE
of polarons.

In conclusion, Hall-effect data on La$_{0.7}$Ca$_{0.3}$MnO$_3$ single
crystal samples reveal clear evidence for the boundary between a
small-lattice-polaron picture, valid at temperatures greater than 1.4$T_C,$
and more complex behavior signalling the onset of metallic behavior. In a
double exchange ferromagnet above its Curie temperature, local ferromagnetic
correlations are enabled by carrier hopping. Localization of the carrier
reduces those correlations, lowering the magnetic free energy mainly through
an increase in spin entropy \cite{varma}. In the presence of a tendency for
lattice-polaron formation, the entropy term tips the free energy balance
toward polaron collapse\cite{emin,LP}. Once small polarons form, the
underlying antiferromagnetic interactions favor local spin arrangements that
suppress Pancharatnam-phase contributions to the Hall effect, leaving
essentially pure polaronic behavior. Therefore, the deviation of the Hall
resistivity from the polaron term Eq.~(\ref{eq-RH}) and the associated
reduction of the effective activation energy for ordinary conductivity mark
the limit of small-polaron behavior. Below that temperature, the system is a
composite of a small-polaron background increasingly filled by ferromagnetic
fluctuations that are more highly conducting and that have a Hall
resistivity dominated by local spin texture. The insulator-metal transition
then proceeds by a thermally-driven percolation transition.

This work was supported in part by DOE DEFG-91ER45439. We acknowledge
helpful discussions with Y. Lyanda-Geller and P. Goldbart.

%TCIMACRO{
%\TeXButton{Figure 1}{\begin{figure}
%\noindent
%\caption{Hall resistivity $\rho_{xy}$ of a ${\rm La_{0.7}Ca_{0.3}MnO_3}$ single crystal 
%as a function of
%field at indicated temperatures (main panel), and 
%the temperature dependence of longitudinal resistivity $\rho_{xx}$ at 0 T and 7 T (inset).}
%\label{fig1}
%\end{figure}%
%}}%
%BeginExpansion
\begin{figure}
\noindent
\caption{Hall resistivity $\rho_{xy}$ of a ${\rm La_{0.7}Ca_{0.3}MnO_3}$ single crystal 
as a function of
field at indicated temperatures (main panel), and 
the temperature dependence of longitudinal resistivity $\rho_{xx}$ at 0 T and 7 T (inset).}
\label{fig1}
\end{figure}%
%
%EndExpansion

%TCIMACRO{
%\TeXButton{Figure 2}{\noindent
%\begin{figure}
%\caption{The activated behavior of Hall coefficient $R_H$ (main panel) and
%Hall mobility $\mu_H$ (inset) above $T_C$. Above 300 K, the slopes are in
%accord with small-polaron hopping theory. Also shown in the main panel is the effective 
%activation energy of the conductivity, $E_{\sigma}^*$.}
%\label{fig2}
%\end{figure}%
%}}%
%BeginExpansion
\noindent
\begin{figure}
\caption{The activated behavior of Hall coefficient $R_H$ (main panel) and
Hall mobility $\mu_H$ (inset) above $T_C$. Above 300 K, the slopes are in
accord with small-polaron hopping theory. Also shown in the main panel is the effective 
activation energy of the conductivity, $E_{\sigma}^*$.}
\label{fig2}
\end{figure}%
%
%EndExpansion

%TCIMACRO{
%\TeXButton{Figure 3}{\begin{figure}
%\caption{The anomalous Hall coefficient $R_S$ (symbols) 
%are compared with $ \rho_{xx}$ (solid line) and $\rho_{xx}^2$ (dotted line). 
%The main panel is
%in a logarithmic scale and the inset is a linear scale plot.}
%\label{fig3}
%\end{figure}%
%}}%
%BeginExpansion
\begin{figure}
\caption{The anomalous Hall coefficient $R_S$ (symbols) 
are compared with $ \rho_{xx}$ (solid line) and $\rho_{xx}^2$ (dotted line). 
The main panel is
in a logarithmic scale and the inset is a linear scale plot.}
\label{fig3}
\end{figure}%
%
%EndExpansion

\end{document}